\documentclass[aps,prb,twocolumn,superscriptaddress]{revtex4}
\usepackage{graphicx}
\usepackage{latexsym}
\usepackage{amssymb}
\usepackage{amsmath}
\usepackage{amsfonts}
\usepackage{bm}
\usepackage{multirow}
\usepackage{color}
\newcommand{\ii}{\mathrm{i}}
\newcommand{\dd}{\mathrm{d}}
\newcommand{\bra}[1]{\langle #1|}
\newcommand{\ket}[1]{|#1 \rangle}

\newcommand{\dsE}{\mathbb{E}}
\newcommand{\scG}{\mathcal{G}}
\newcommand{\scV}{\mathcal{V}}
\newcommand{\scE}{\mathcal{E}}
\newcommand{\Tr}{\mathop{\mathrm{Tr}}}
\newcommand{\pf}{\mathop{\mathrm{pf}}}

\newcommand{\eq}[1]{\begin{equation}#1\end{equation}}
\newcommand{\eqs}[1]{\begin{equation}\begin{split}#1\end{split}\end{equation}}

\newcommand{\eqnref}[1]{Eq.\,\eqref{#1}}
\newcommand{\figref}[1]{Fig.\,\ref{#1}}
\newcommand{\tabref}[1]{Tab.\,\ref{#1}}

\newcommand{\dia}[3]{\raisebox{#3pt}{\includegraphics[height=#2pt]{dia_#1}}}

\definecolor{pink1}{rgb}{0.858, 0.188, 0.478}

\begin{document}

\title{Machine Learning Spatial Geometry from Entanglement Features}
\author{Yi-Zhuang You}
\affiliation{Department of physics, Harvard University, Cambridge MA 02138, USA}
\author{Zhao Yang}
\affiliation{Department of physics, Stanford University, CA 94305, USA}
\author{Xiao-Liang Qi}
\affiliation{Department of physics, Stanford University, CA 94305, USA}
\date{\today}
\begin{abstract}
Motivated by the close relations of the renormalization group with both the holography duality and the deep learning, we propose that the holographic geometry can emerge from deep learning the entanglement feature of a quantum many-body state. We develop a concrete algorithm, call the entanglement feature learning (EFL), based on the random tensor network (RTN) model for the tensor network holography. We show that each RTN can be mapped to a Boltzmann machine, trained by the entanglement entropies over all subregions of a given quantum many-body state. The goal is to construct the optimal RTN that best reproduce the entanglement feature. The RTN geometry can then be interpreted as the emergent holographic geometry. We demonstrate the EFL algorithm on 1D free fermion system and observe the emergence of the hyperbolic geometry (AdS$_3$ spatial geometry) as we tune the fermion system towards the gapless critical point (CFT$_2$ point).
\end{abstract}
\maketitle

\section{Introduction}

Holographic duality\cite{Witten:1998ty,Gubser:1998fe,Maldacena:1999fk} is a duality proposed between a quantum field theory (the boundary theory) and a gravitational theory (the bulk theory) in one higher dimension. 
In 2006, S. Ryu and T. Takayanagi proposed the Ryu-Takayanagi (RT) formula\cite{Ryu:2006fj}, which relates the entanglement entropy of a boundary region to the area of the minimal surface in the bulk that is homologous to the same region. The RT formula and its generalizations\cite{Hubeny:2007cf,faulkner2013quantum,dong2014holographic,dong2016gravity} point out that entanglement plays a fundamental rule in holographic duality. One perspective to understand the entanglement-geometry correspondence is to consider a tensor network representation of a quantum many-body state\cite{Vidal:2007pi,Vidal:2008zp}, and view the network geometry as a representation of the dual spatial geometry\cite{Swingle:2012bs,Swingle:2012yq}. Many different schemes of tensor network approaches have been investigated\cite{Vidal:2007pi,Vidal:2008zp,Evenbly:2009hb,Takayanagi:2012vm,Verstraete:2013do,Takayanagi:2014an,Molina-Vilaplana:2015lr,Takayanagi:2015pd,Wen:2016dz,Qi:2013fm,Qi:2015ct,Gu:2016ud,Lee:2014qh,Lee:2015aa,You:2016wq,Levin:2007sj,Evenbly:2015kl,Evenbly:2015yg}. Tensor network states with various entanglement properties similar to holographic theories have been constructed\cite{Preskill:2015ja,Yang:2016fu,Hayden:2016zm,Qi:2017qf,donnelly2017living,han2017discrete,may2017tensor,Hyatt:2017df}. In particular, the random tensor network (RTN) states \cite{Hayden:2016zm} are shown to satisfy the Ryu-Takayanagi formula\cite{Ryu:2006fj} and the quantum error correction properties\cite{Almheiri:2015ei} in the large bond dimension limit. The RTN states on all possible graphs form an overcomplete basis of the boundary Hilbert space\cite{Qi:2017qf}, so that a generic many-body state of the boundary can be mapped to a superposition of RTN's with different geometry. For states with a semi-classical bulk dual, one expects the superposition to be strongly peaked around a ``classical geometry", which provides a best approximation to entanglement entropy of different regions in the given state. In other words, finding the best RTN description of a given many-body state can be considered as a variational problem similar to a familiar variational wavefunction approach, except that the criteria of the optimization is not minimizing energy but reproducing {\it entanglement features} of the state, such as entanglement entropy and Renyi entropies of various subsystems. For deeper understanding of holographic duality, such as understanding how boundary dynamics are mapped to bulk gravitational dynamics, it is essential to develop a systematic approach of finding the optimal network geometry for generic many-body states.

\begin{figure}[htbp]
\begin{center}
\includegraphics[width=0.34\textwidth]{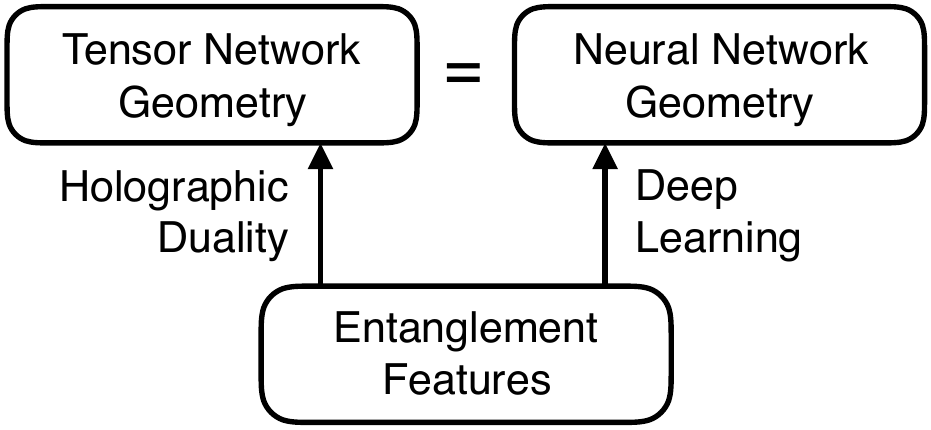}
\caption{Conceptual connections between holographic duality and deep learning.}
\label{fig: concepts}
\end{center}
\end{figure}

In this paper, we propose that the RTN optimization problem can be mapped to a deep learning problem\cite{Hinton:2006tw,Bengio:2007jl,LeCun:2015cr}, because the paradigm of neural network based deep learning is precisely about how to adjust the network connectivity (geometry) to achieve a certain optimization goal. More specifically, we propose a  learning approach, called the \emph{entanglement feature learning} (EFL), which learns the entanglement features in the quantum many-body state and encodes the entanglement structures in the neural network connectivity. Interestingly, the deep learning approach provides not only a technical tool to optimize the RTN, but also a profound connection between tensor networks and neural networks in terms of their geometric interpretations. Base on this interpretation, the holographic dual spatial geometry of a quantum many-body state could emerge as the neural network geometry from machine learning the entanglement features. {\it In other words, spacial geometry is just an efficient way to encode entanglement features.} The corresponding tensor network can be viewed as a disentangling circuit that gradually resolves the entanglement features at different layers, which is the common idea underlying other tensor network holography approaches.\cite{Hyatt:2017df} For simplicity we will consider the second Renyi entropy $S^{(2)}(A)$ of all subregions $A$ as entanglement features of a state. Using the second Renyi entropy of some regions as the training data, the goal of the neural network is to give a best prediction to the second Renyi entropy of other regions. As the learning is done, the geometric structure of the neural network can be interpreted as the emergent holographic bulk geometry. This draws a direct connection between holographic duality and the deep learning, as illustrated in \figref{fig: concepts}. This connection was also made in a recent work Ref.\,\onlinecite{Gan:2017xy}, based on the similarity in  their relations to the renormalization group\cite{Verlinde:2000pn,Evenbly:2009hb,Swingle:2012bs,Swingle:2012yq,Takayanagi:2012vm,Balasubramanian:2013xz,Verstraete:2013do,Beny:2013ch,Mehta:2014qf,Lin:2016km}. The relation between neuron networks and tensor networks have also been discussed recently in Refs.\,\onlinecite{Miles:2016hd,Chen:2017ed,Gao:2017ml,Huang:2017nb,Carleo:2017jv,Deng:2017mw}.

\begin{table}[htp]
\caption{A terminology dictionary of EFL}
\begin{center}
\begin{tabular}{l|l}
machine learning & EFL\\
\hline
neural network & random tensor network\\
visible units & boundary tensors\\
hidden units & bulk tensors \\
edge weight & edge mutual information \\
training samples & entanglement regions \\
input data  & entanglement feature \\
network geometry & bulk spatial geometry
\end{tabular}
\end{center}
\label{tab: dictionary}
\end{table}

In this work, we point out that for ``machine learning holography'', what should be learned are the entanglement features of the quantum many-body states. We also develop a concrete EFL algorithm that can be implemented in numerics. A terminology dictionary of EFL is summarized in \tabref{tab: dictionary}. Our EFL approach is based on a deep learning architecture known as the deep Boltzmann machine (DBM)\cite{Salakhutdinov:2009db,Norouzi:2009rz,Hu:2016fv}. Boltzmann machines are a class of machine learning models that has been introduced condensed matter physics research in many recent works\cite{Deng:2016fr,Torlai:2016fk,Aoki:2016yg,Chen:2017ed,Gao:2017ml,Huang:2017nb,Huang:2017xy,Deng:2017mw,Weinstein:2017ad,Decelle:2017dk}, in particular Ref.\,\onlinecite{Torlai:2016fk} contains a nice review of Boltzmann Machines for physicists. We show that each RTN can be mapped to a DBM with the same network structure, therefore the optimal RTN state can be found by training the corresponding DBM.  However, there is no efficient method to train a generic DBM, so we have to make some restrictions to the neural network architecture in order to make EFL a practical (rather than theoretical) algorithm. To this end, we will restrict to the RTN on planar graphs. It turns out that the planar RTN already has sufficient expression power to represent a rich variety of states from area law\cite{Srednicki:1993rv,Verstraete:2006qt,Hastings:2007sd} to volume-law\cite{Page:1993fv,Foong:1994bf,Sen:1996rw} entanglement. We develop an efficient deterministic learning approach based on the exact solution of planar graph Ising models. We then demonstrate the EFL in 1D free fermion systems and show how the holographic geometry grows deeper in the perpendicular direction as the boundary fermion state approaches critical point.

The remainder of this paper is organized as follows. In Section \ref{sec: RTN theory}, we will first review the construction of the RTN and its entanglement properties. In Section\ref{sec: EFL}, we will propose the EFL algorithm and analyze some of the technical challenges. In Section \ref{sec: numerics}, we will apply the EFL on a 1D free fermion model and demonstrate how the holographic bulk geometry can arise from learning the entanglement features. 

\section{Random Tensor Networks}\label{sec: RTN theory}
\subsection{Definition of RTN States}
\begin{figure}[htbp]
\begin{center}
\includegraphics[width=0.22\textwidth]{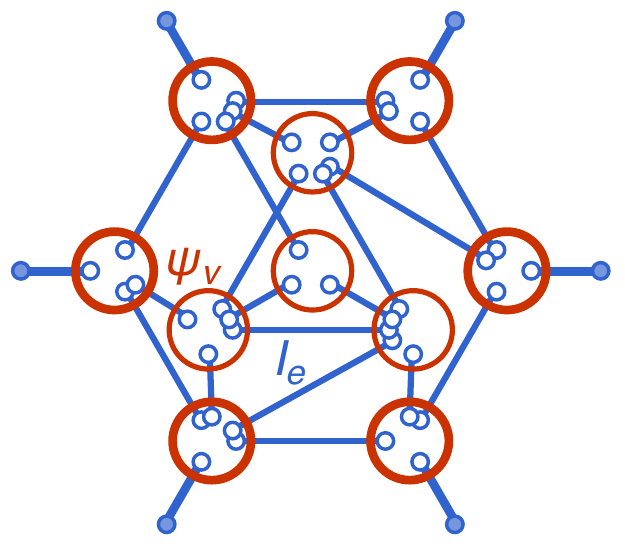}
\caption{The structure of a RTN state.}
\label{fig: RTN}
\end{center}
\end{figure}

We will briefly review the definition of random tensor network (RTN) state following the projected entangled pair state (PEPS) approach.\cite{Hayden:2016zm,Qi:2017qf} A RTN state is specified by an edge-weighted graph $\scG=(\scV;\scE,I)$ comprising the vertex set $\scV$ and the edge set $\scE$ along with a weighting function $I:\scE\to\mathbb{R}_+$, such that each edge $e\in\scE$ is associated with a real and positive edge weight $I_e$. On each vertex $v\in \scV$, we define a local Hilbert space $\mathcal{H}_v=\bigotimes_{e\in\dd v}\mathcal{H}_{v}^{e}$, where $\dd v$ denotes the set of edges adjacent to the vertex $v$. $\mathcal{H}_{v}^{e}$ is subspace on the vertex $v$ to be connected to the incident edge $e$ (as the small blue circle in \figref{fig: RTN}). Let $\ket{\mu_v^e}$ (labeled by $\mu_v^e=1,2,...$) be a complete set of basis states of the Hilbert space $\mathcal{H}_{v}^{e}$.

We then define a random state $\ket{\psi_v}\in\mathcal{H}_v$ on each vertex $v$ (as the big red circle in \figref{fig: RTN}).
\eq{\ket{\psi_v}=\sum_{[\mu_{v}]}T[\mu_{v}]\bigotimes_{e\in\dd v}\ket{\mu_{v}^{e}}.}
The coefficient tensor $T$ is a random tensor, whose tensor elements are independently drawn from normal distributions following $P(T)\propto e^{-\frac{1}{2}\sum_{[\mu_{v}]}|T[\mu_{v}]|^2}$. On each edge, we define an entangled pair state $\ket{I_e}$ (as the blue link in \figref{fig: RTN}) in the Hilbert space $\bigotimes_{v\in\partial e}\mathcal{H}_{v}^{e}$ (where $\partial e$ denotes the set of two vertices at the end of the edge $e$),
\eq{\ket{I_e}=\sum_{[\mu^{e}]}\lambda[\mu^{e}]\bigotimes_{v\in\partial e}\ket{\mu_{v}^{e}}.}
The entanglement of $\ket{I_e}$ across the edge is characterized by the edge mutual information $I_e$. Each edge could have a different $I_e$ in general. If we treat the coefficient $\lambda[\mu^{e}]=\lambda_{\mu^{e}_1\mu^{e}_2}$ as a matrix, the $n$th Renyi mutual information can be expressed as
\eq{I_e^{(n)}=\frac{2}{1-n}\ln\Tr(\lambda\lambda^\dagger)^n.}
In the following, we will focus on the case of Renyi index $n=2$ and take $I_e=I_e^{(2)}$ unless otherwise specified. It is free to choose $\lambda$ on each edge, as long as the edge mutual information $I_e$ matches the edge weight $I_e$ of the graph $\scG$. There is also a set of special edges (the thick edges in \figref{fig: RTN}) on the boundary of the network. They are the external edges (physical legs) that connects to the physical degrees of freedom. On these edges, we assume that the entangled pair states are maximally entangled, hence the edge mutual information is $2\ln D_\partial$ with $D_\partial$ being the bond dimension of the external leg.

Given the random state $\ket{\psi_v}$ on each vertex $v$ and the entangled pair state $\ket{I_e}$ on each edge $e$, the RTN state can be constructed by projecting the entangled pair states to random vertex states via the following partial inner product
\eq{\ket{\scG}=\bigotimes_{v\in\scV}\bigotimes_{e\in\scE}\langle\psi_v|I_e\rangle.}
The remaining subspaces (as solid circles in \figref{fig: RTN}) on the dangling ends of the external edges are not touched by the projection. They form the physical Hilbert space $\mathcal{H}^\text{phy}=\bigotimes_{v\in\scV_\partial}\mathcal{H}_v^\text{phy}$ in which the RTN state $\ket{\scG}$ is supported. Here $\scV_\partial$ denotes the set of boundary vertices, i.e.\,the subset of $\scV$ whose vertices are connected to the external edges. It is worth mentioning that $\ket{\scG}$ should better be treated as an \emph{ensemble} of RTN states, instead of a single specific state, due to the randomness in $\ket{\psi_v}$. All states in the ensemble are labeled by the same edge-weighted graph $\scG$ and share the similar entanglement feature.

\subsection{Entanglement Features of RTN States}

The \emph{entanglement feature} of a quantum many-body state refers to the full set of entanglement entropies over all entanglement subregions. In general, one could include all orders of Renyi entropies in the definition, but we will only focus on the 2nd Renyi entropies in the following and leave the generic discussion to the last section.

Given an ensemble of RTN states $\ket{\scG}$ and a subregion $A\subseteq\scV_\partial$, the ensemble-typical 2nd Renyi entropy $S_{\scG}(A)$ over the subregion $A$ is defined via
\eq{\label{eq: S def}
e^{-S_{\scG}(A)}=\dsE\frac{\Tr_{A}(\Tr_{\bar{A}}\ket{\scG}\bra{\scG})^2}{(\Tr\ket{\scG}\bra{\scG})^2},}
where $\dsE$ takes the RTN ensemble expectation value (i.e.\,averaging over the random states $\ket{\psi_v}$ on all vertices), and $\bar{A}=\scV_\partial\setminus A$ denotes the complement region of $A$. We have explicitly introduced the denominator $\Tr\ket{\scG}\bra{\scG}$ to ensure the normalization of the RTN density matrix. An important result of Ref.\,\onlinecite{Hayden:2016zm} is to show that the entanglement entropy $S_\scG(A)$ can be expressed in term of the free energies of a classical Ising model on the same graph $\scG$ in the large bond dimension limit. A more general treatment away from that limit is provided in a related work Ref.\,\onlinecite{Vasseur:2017om}, but in this work we will only consider the large bond dimension limit.

\begin{figure}[htbp]
\begin{center}
\includegraphics[width=0.43\textwidth]{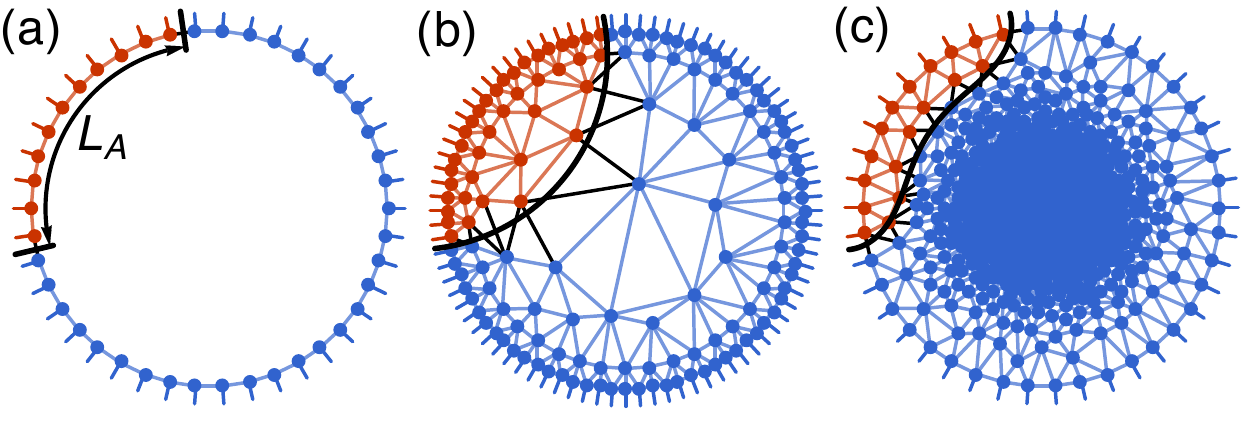}
\caption{Entanglement entropy as the minimal cut (in black) through the tensor network that separates the region $A$ (in red) from $\bar{A}$ (in blue). The Ising domain wall is automatically the minimal cut in the large bond dimension (low temperature) limit. Different network structures gives rise to different scaling behaviors of the entanglement entropy: (a) area law $S(A)\sim\text{const.}$, (b) logarithmic law $S(A)\sim\ln L_A$, (c) volume law $S(A)\sim L_A$.}
\label{fig: laws}
\end{center}
\end{figure}

To specify the Ising model, we first introduce a set of Ising spins $\sigma_v=\pm1$ for all $v\in\scV$ and an additional set of Ising variables $\tau_v=\pm1$ on the boundary $v\in \scV_\partial$ only. The model is described by the energy functional
\eq{\label{eq: E Ising}
E_\scG[\sigma,\tau]=-\sum_{e\in\scE}J_e\prod_{v\in\partial e}\sigma_v-h\sum_{v\in\scV_\partial}\tau_v\sigma_v.}
The Ising coupling $J_e\equiv I_e/4$ is set by the edge mutual information $I_e$ of the RTN state. The external field $h\equiv\frac{1}{2}\ln D_\partial$ is set by the local Hilbert space dimension $D_\partial$ of the physical degrees of freedom (which is also the bond dimension of the external leg). Only $\sigma_v$ spins are dynamical, and $\tau_v$ are just Ising variables that specifies the directions of the external pinning field $h\tau_v$ on the boundary. The configuration of $\tau_v$ is determined by the choice of the entanglement region $A$
\eq{\label{eq: tau A}\tau_v(A)=\left\{\begin{array}{ll}-1 & v\in A, \\ +1 & v\in \bar{A}.\end{array}\right.}
Tracing out the dynamical spins $\sigma_v$, the free energy $F[\tau]$ of the boundary spins $\tau_v$ can be  defined via
\eq{e^{-F_\scG[\tau]}=\sum_{[\sigma]}e^{-E_\scG[\sigma,\tau]}.}
In the large bond dimension limit ($I_e\gg1$), it was shown\cite{Hayden:2016zm} that the typical 2nd Renyi entropy of the RTN state $\ket{\scG}$ is given by the free energy difference
\eq{\label{eq: S RTN}S_\scG(A)=F_\scG[\tau(A)]-F_\scG[\tau(\emptyset)],}
where $\tau(A)$ denotes the boundary pinning field configuration specified in \eqnref{eq: tau A} and $\tau(\emptyset)$ denotes the configuration of $\tau_v=+1$ for all $v\in\scV_\partial$. The derivation of \eqnref{eq: S RTN} is reviewed in Appendix \ref{sec: RTN}. The physical intuition of \eqnref{eq: S RTN} comes from the interpretation\cite{Ryu:2006fj} of the entanglement entropy as the area of the minimal surface that separates the region $A$ from $\bar{A}$ in the holographic bulk. Correspondingly, the free energy difference $F[\tau(A)]-F[\tau(\emptyset)]$ measures the energy cost of the domain wall that separates the part $A$ from $\bar{A}$ in the tensor network (see \figref{fig: laws}), which matches the holographic interpretation of the entanglement entropy in the large bond dimension limit. Technically, the advantage of RTN over other types of tensor networks also lies in the fact that the 2nd Renyi entropy of the RTN state can be efficiently estimated from the free energy of the corresponding Ising model as in \eqnref{eq: S RTN}. For a generic tensor network, calculating its entanglement entropy requires to diagonalize the reduced density matrix, which could be much more difficult than solving the Ising model in many cases.

The set of entanglement entropies $\{S_\scG(A)|A\subseteq\scV_\partial\}$ constitutes the entanglement feature of the RTN state, which only depends on the graph $\scG$ and its edge weights $I_e$. The RTN state thus provides us a model to encode the entanglement feature directly in the network structure ({\it i.e.} the graph geometry). This is the essential idea behind the tensor network holography. In many previous approaches, a bulk geometry is first given and a tensor network is tiled on the background geometry. The resulting tensor network state then produces the entanglement feature on the holographic boundary that is dual to the holographic bulk geometry. For example, \figref{fig: laws} demonstrates how different network structures lead to different scaling behaviors of the single-interval entanglement entropy. However, in this work, we would like to consider the inverse problem: given the entanglement feature of a quantum many-body state, how to determine the optimal holographic geometry? We will show that this problem can be mapped to a machine learning problem, which we called the \emph{entanglement feature learning} (EFL).

\section{Entanglement Feature Learning}\label{sec: EFL}
\subsection{General Algorithm}
The goal of EFL is to develop an RTN ensemble that best matches the entanglement feature of the given many-body state $\ket{\Psi}$. The graph geometry of the RTN is then interpreted as the dual bulk geometry. In principle, all graph geometries can be realized on a complete graph (the graph with all-to-all connections) by adjusting the edge weights $I_e$. For example, an edge in the complete graph can be disconnected by setting its weight $I_e=0$ to zero. Therefore optimizing the graph geometry is equivalent to optimizing the set of edge weights on the complete graph, and the latter is a typical problem of the neural network based deep learning. We will apply the deep learning technique to optimize the random tensor network connections and obtain the optimal holographic geometry of the given quantum many-body state. 

Given a quantum many-body state $\ket{\Psi}$ (to learn), we first extract its entanglement feature by collecting the 2nd Renyi entanglement entropies $S_\Psi(A)$ over different entanglement subregions $A$
\eq{S_\Psi(A)=-\ln\Tr_A(\Tr_{\bar{A}}\ket{\Psi}\bra{\Psi})^2.}
Admittedly, calculating the entanglement entropy of a generic many-body state is difficult. However, let us assume that these data can be in principle collected, for example by experimental measurements\cite{Islam:2015cs,Garttner:2016ss,Li:2017qq,Pichler:2016dw}. Then they can be used to construct the training set:
\eq{\{(\tau(A),S_\Psi(A))|A\subseteq\scV_\partial\},}
where $\tau(A)$ is the boundary pinning field configuration defined in \eqnref{eq: tau A}, which is just another way to specify the entanglement region $A$. Usually it is not practical to collect  entanglement entropies for all possible subregions $A\subseteq\scV_\partial$, so only a subset of the entanglement feature will be used in the EFL (how to sample the subset will be explained in details later). Once the entanglement feature is collected, we will make no further reference to the original quantum state $\ket{\Psi}$.

We wish to fit the entanglement feature of the given state $\ket{\Psi}$ by the RTN state $\ket{\scG}$. We would like to emphasize that we are not intended to find the tensor network representation of the state $\ket{\Psi}$, which could be a much harder task. We just want to find the optimal tensor network geometry such that the entanglement features between $\ket{\Psi}$ and $\ket{\scG}$ match as much as possible. In fact, as the tensors are random in the RTN, the RTN state $\ket{\scG}$ would be very different from (most likely orthogonal to) the given state $\ket{\Psi}$. To learn the tensor network geometry from the entanglement feature, there are two possible learning approaches: \emph{supervised} learning or \emph{unsupervised} learning. 

In supervised learning, each training sample is a pair $(\tau(A),S_\Psi(A))$ consisting of the Ising configuration $\tau(A)$ as the input object and the entanglement entropy $S_\Psi(A)$ as the desired output value. The supervised learning will seek for a fitting function $S_\scG(A)$ based on the RTN model with minimal prediction error. The supervised EFL is essentially a regression problem. We can choose to minimize the mean square error loss function, which is commonly used for regression problems
\eq{\label{eq: L supervised}\mathcal{L}(\scG)=\mathop{\text{avg}}_{A\subseteq \scV_\partial}(S_\scG(A)-S_\Psi(A))^2.}
The variational parameters will be the edge weights $I_e$ that parameterize the graph $\scG$ (and the RTN model).

In unsupervised learning, the training samples $\tau(A)$ are ``unlabeled'', but they appear with an empirical probability distribution
\eq{\label{eq: P Psi}P_\Psi[\tau(A)]\propto e^{-S_\Psi(A)}.}
Such training set can be prepared by Monte Carlo sampling the entanglement region $A$ following the Boltzmann weight $e^{-S_\Psi(A)}$ on the given state $\ket{\Psi}$. The goal of the unsupervised learning is to train a generative model that could reproduce the samples $\tau(A)$ with the probability distribution close to the empirical distribution as much as possible. If the goal is achieved, it is believed that the generative model has capture the hidden features in the training data. For our purpose, we take the RTN model as the generative model, which generates the sample $\tau(A)$ with the model probability
\eq{P_\scG[\tau(A)]\propto e^{-S_\scG(A)}\propto e^{-F_\scG[\tau(A)]},}
or more precisely,
\eqs{P_\scG[\tau]&=\frac{1}{Z_\scG}\sum_{[\sigma]}e^{-E_\scG[\sigma,\tau]},\\
Z_\scG&=\sum_{[\tau]}e^{-F_\scG[\tau]}=\sum_{[\sigma,\tau]}e^{-E_\scG[\sigma,\tau]},}
where the energy model $E_\scG[\sigma,\tau]$ is given by \eqnref{eq: E Ising}. If we treat the bulk spins $\sigma$ as hidden units and the boundary spins $\tau$ as visible units, the model is precisely mapped to the Boltzmann machine\cite{Hinton:1986fk,Aarts:1988nr} in machine learning. The goal is to approximate the empirical distribution $P_\Psi[\tau]$ by the distribution $P_\scG[\tau]$ produced by the Boltzmann machine. To measure how similar the two distributions are, the Kullback-Leibler divergence is typically used as the objective function
\eq{\mathcal{L}(\scG)=\sum_{[\tau]}P_\Psi[\tau]\ln \frac{P_\Psi[\tau]}{P_\scG[\tau]},}
which is minimized if $P_\scG[\tau]\to P_\Psi[\tau]$. Because the empirical distribution $P_\Psi[\tau]$ was constructed in \eqnref{eq: P Psi} to encode the entanglement feature of $\ket{\Psi}$, if the Boltzmann machine managed to reproduce this distribution after training, the entanglement feature should have been learnt and encoded in the neural network connectivity, which gives us a representation of the emergent holographic bulk geometry.

\begin{figure}[htbp]
\begin{center}
\includegraphics[width=0.22\textwidth]{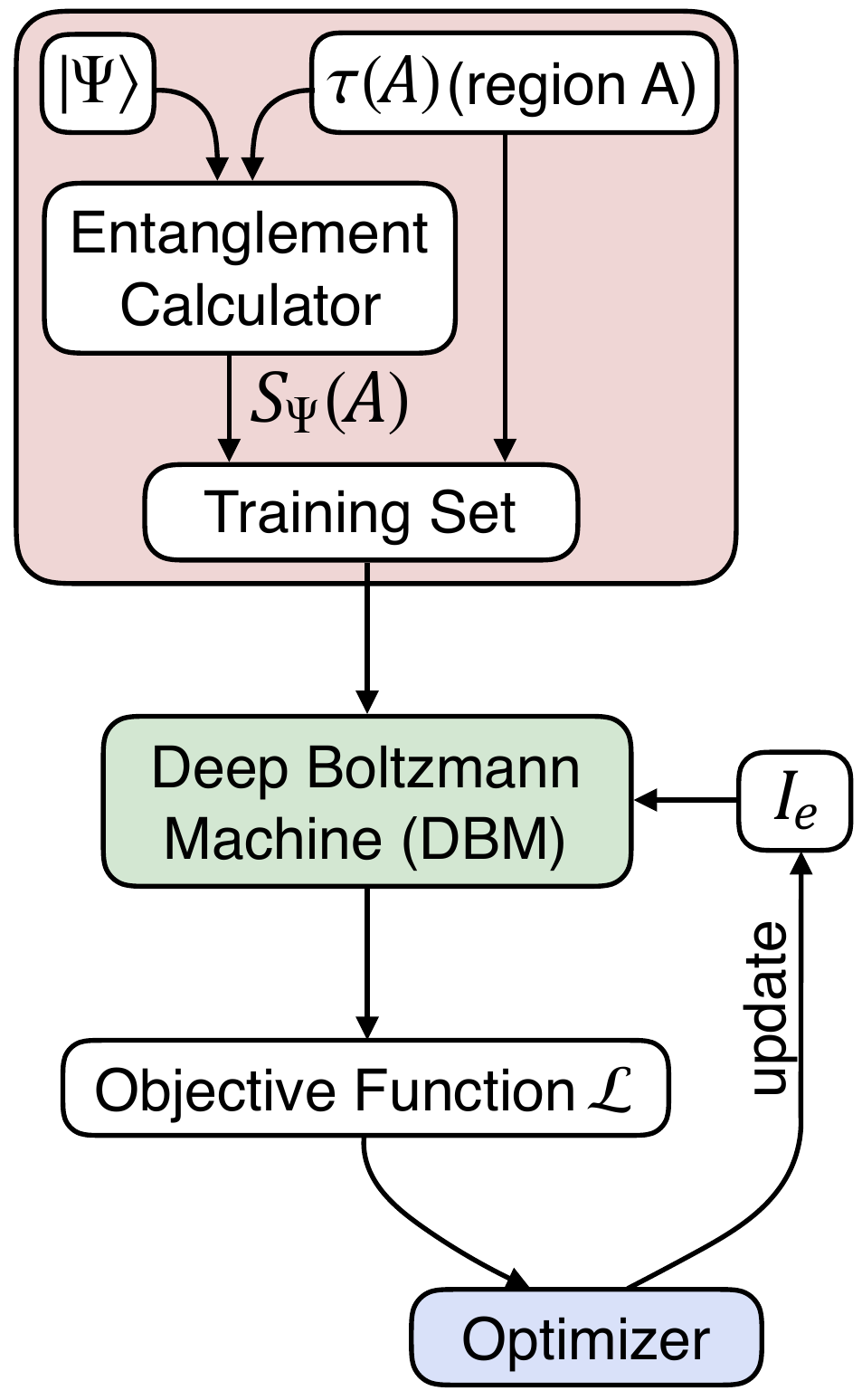}
\caption{Computation graph of EFL. Arrows indicate the directions that the data flows. The training data is prepared in the red module.}
\label{fig: graph}
\end{center}
\end{figure}

For both supervised and unsupervised learning, the training procedure is to minimize objective function $\mathcal{L}(\scG)$, which is formally a functional of the edge-weighted graph $\scG$. As mentioned before, we can always embed the graph $\scG$ in a large enough complete graph and take the edge weights $I_e$ (or equivalently the Ising couplings $J_e=I_e/4$) as the variational parameters. Hence, we can use a gradient descent algorithm over $\mathcal{L}(\scG)$ to find its minimum according to the following update rule
\eq{I_e\leftarrow I_e-r_l\frac{\partial\mathcal{L}(\scG)}{\partial I_e},}
where $r_l$ denotes the learning rate. The whole EFL algorithm is summarized as the computation graph in \figref{fig: graph}. In the training process, the neural network learns the entanglement feature of the input quantum state $\ket{\Psi}$. As the training converges, we open up the neural network and extract the network connectivity from the parameters $I_e$, which parameterize the optimal edge mutual information of the RTN as well as the optimal graph geometry in the holographic bulk. 

In this work, we will adopt the supervised learning approach and leave the unsupervised learning approach for future investigation. There are still two technical obstacles that we have to overcome to make the EFL really a practical (rather than theoretical) algorithm for tensor network holography. In the remainder of this section, we will analyze the obstacles and provide solutions to them.

\subsection{Deterministic Learning on Planar Graph}

The gradient descent method is not practical for training generic Boltzmann machines with unrestricted connections (i.e.\,on a complete graph). One major reason lies in the lack of efficient inference method: the complexity to evaluate the free energy $F_\scG[\tau]$ (or the marginal distribution $P_\scG[\tau]$) of the Boltzmann machine grows exponentially with the number of hidden units. Adding restrictions to the network structure allows for more efficient training algorithms, such as the restricted Boltzmann machine (RBM).\cite{Smolensky:1986hl,Hinton:2006tw,Hinton:2010qq} By stacking RBM layers, one obtains a deep architecture known as the deep Boltzmann machine (DBM),\cite{Salakhutdinov:2009db} which better fits the purpose of EFL to produce the geometry deep in the holographic bulk. However, the original proposal to estimate the learning gradient for the DBM is based on the Monte Carlo method. It is found that the deep layers typically receive very weak gradient signals, which can be easily overwhelmed by the thermal fluctuations introduced by the Monte Carlo process. The net effect is that the thermal noise will drive the edge weights $I_e$ in the deep layers to follow a random walk until the activation saturates. Therefore the deep layers can not be trained by stochastic learning algorithms. Instead, we need \emph{deterministic} learning\cite{Decelle:2017dk} algorithms. The idea is to avoid the Monte Carlo sampling and evaluate the Ising model free energy analytically. Several approximate methods have been developed, including the belief propagation\cite{Mezard:2002ij,Yedidia:2003dp,Tramel:2016ud} and the high-temperature series expansion.\cite{Gabrie:2015fv} 

Instead of approximate approaches, if we restrict the network geometry to planar graphs, there are exact learning methods\cite{Globerson:2007zr,Schraudolph:2008qp,Johnson:2010tx}, exploiting the exact solvability of planar graph Ising models\cite{Kac:1952qr,Fisher:1961il,Kasteleyn:1961bl} (by mapping them to free Majorana fermion problems on related but different graphs, see Appendix \ref{sec: Ising} for details). Naively, it seems too restricted to study planar graphs, which are very special among all graphs. However, the RTN on a planar graph can already model a variety of entanglement features on the holographic boundary, as demonstrated in \figref{fig: laws}. For example, the volume-law entanglement can be described by a planar network geometry with flat or positive curvature, because in that case the minimal surface is pushed to the boundary. Therefore the planar graphs can describe a large family of states of interest,\footnote{There are certainly states that cannot be described by a planar RTN. The simplest example is a state with two long-range EPR pairs between boundary points $x_1,x_3$ and $x_2,x_4$, with the points ordered as $x_1<x_2<x_3<x_4$. } including, for example, area-law\cite{Srednicki:1993rv,Verstraete:2006qt,Hastings:2007sd}  ground states of local Hamiltonians and volume-law\cite{Page:1993fv,Foong:1994bf,Sen:1996rw} excited eigenstates satisfying the eigenstate thermalization hypothesis\cite{Deutsch:1991ik,Srednicki:1994ns,Tasaki:1998vy}.

Details of the Ising-Majorana fermion mapping is reviewed in Appendix \ref{sec: Ising}. As a short summary, the free energy $F_\scG[\tau]$ can be calculated from the Pfaffian of the lattice adjacency matrix $A[J,h\tau]$ (with edges weighted by $J$ and $h\tau$) on which the dual fermions live:
\eq{F_\scG[\tau]=\sum_{e\in\scE}J_e+h\sum_{v\in\scV_\partial}\tau_v-\ln\pf A[J,h\tau].}
The computational complexity is of the cubic order of the graph size. The gradient can also be calculated efficiently from $\dd(\ln\pf A)=\frac{1}{2}\Tr A^{-1}\dd A$. Because there is no thermal fluctuation in the gradient signal, the edge weights in deep layers can be trained towards their optimal values deterministically. On the other hand, considering the DBM with planar graph architecture is also physically plausible for the purpose of the tensor network holography, because the planar graph is naturally a discretized description of the 2D spatial part of the $(2+1)$D holographic geometry (as the holographic dual to the $(1+1)$D quantum many-body state). 

\subsection{Architecture and Regularization}

\begin{figure}[htbp]
\begin{center}
\includegraphics[width=0.242\textwidth]{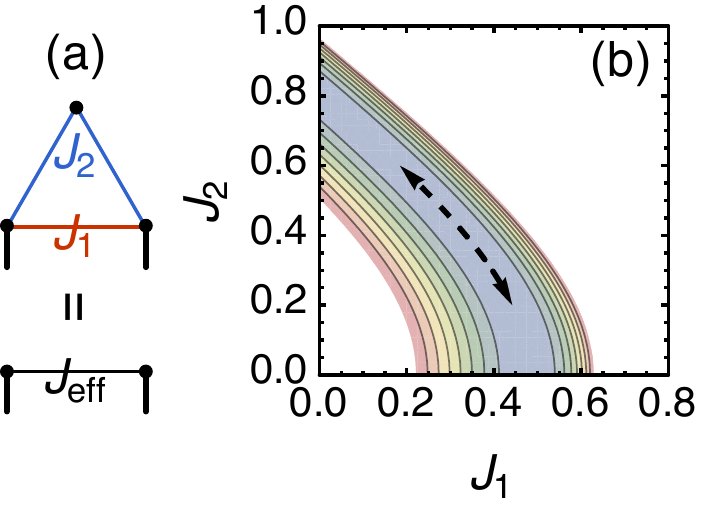}
\caption{Redundancy in a three-vertex graph. (a) The network structure. (b) The objective function in the $J_1$-$J_2$ plane. A flat channel indicates the redundant direction.}
\label{fig: gauge}
\end{center}
\end{figure}

Besides the deterministic learning, another technical challenge of EFL is the redundancy in the graphical representation of the entanglement feature. For example, consider an Ising model with three spins as shown in \figref{fig: gauge}(a), described by
\eq{E[\sigma,\tau]=-J_1\sigma_1\sigma_2-J_2(\sigma_1+\sigma_2)\sigma_3-h(\tau_1\sigma_1+\tau_2\sigma_2),}
which is parameterized by two Ising couplings $J_1$ and $J_2$. But the free energy $F[\tau]=-\ln\sum_{[\sigma]}e^{-E[\sigma,\tau]}$ only depends on an effective coupling (obtained by first tracing out the $\sigma_3$ spin)
\eq{J_\text{eff}=J_1+\frac{1}{2}\ln\cosh 2J_2.}
So there is a trade-off between $J_1$ and $J_2$: as long as $J_\text{eff}$ remains unchanged, adjusting $J_1$ and $J_2$ in the opposite way will not change the free energy $F[\tau]$, and thus will not  affect the objective function. As illustrated in \figref{fig: gauge}(b), there will be a flat channel along which all different edge weights are degenerated in the objective function.

This phenomenon can be viewed as a discrete analog of the diffeomorphism redundancy of the gravity theory. It also poses a problem to the EFL, because each time the training will end up with a different edge weight configuration along the flat direction, which makes it hard to compare the network geometries between two training. Before coming up with a systematic classification of these redundancies, we have to introduce a ``gauge fixing" by hand. This is done by imposing more restrictions on the architecture. 

\begin{figure}[htbp]
\begin{center}
\includegraphics[width=0.44\textwidth]{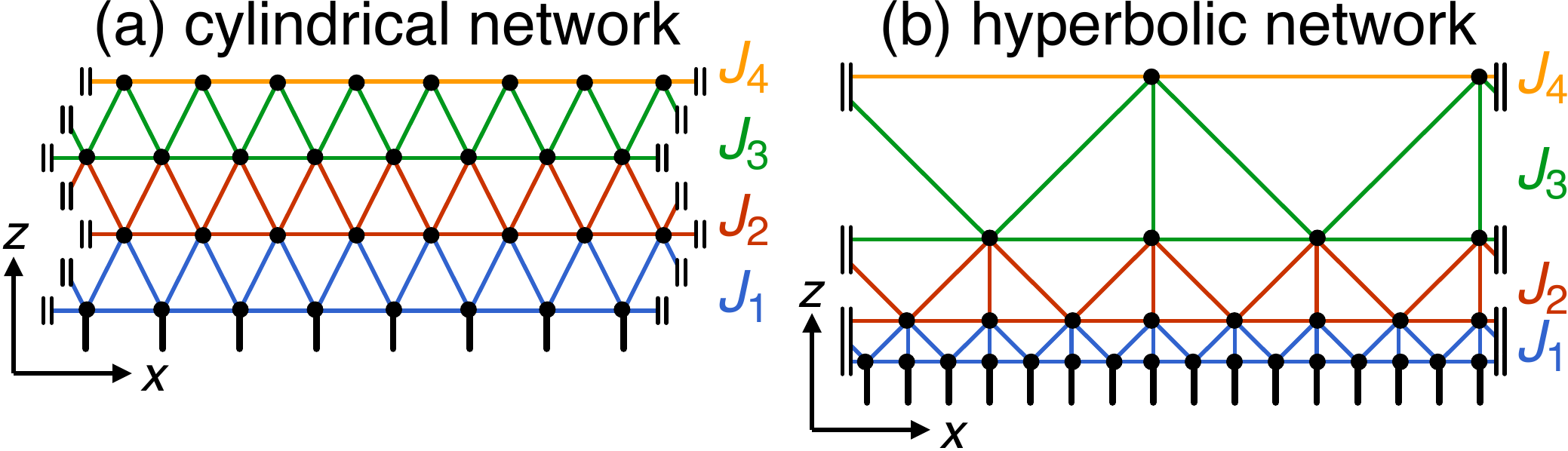}
\caption{Two architectures of planar graph DBM: (a) cylindrical network and (b) hyperbolic network. Both assume periodic boundary condition along the $x$-direction. The bonds of the same color are locked to the same coupling strength $J_z$.}
\label{fig: networks}
\end{center}
\end{figure}

In the following we will consider two particular architectures: the cylindrical and the hyperbolic network as shown in \figref{fig: networks}. In particular, the hyperbolic network in \figref{fig: networks}(b) can be viewed as a variation of the convolutional deep Boltzmann machine architecture. Both networks have layered structure. Within each layer, the horizontal (intra-layer) bonds and the zig-zag (interlayer) bonds can trade off each other (approximately), similar to the situation in \figref{fig: gauge}. To fix this redundancy, we lock the interlayer coupling to the intra-layer coupling on the UV side (i.e.\,the side closer to the boundary), see \figref{fig: networks}. If the training data is translation invariant along $x$-direction, we will also set the coupling uniform within each layer to respect the translation symmetry. 

The learning signal originates from the training data and is passed down layer-by-layer from the boundary into the bulk. Suppose at the beginning, all couplings are initialized to zero. When the training data is presented to the machine, the first layer learns the local spin correlation in the training samples and develops the coupling $J_1$ to match the correlation. Due to the interlayer couplings, the spin correlation in the first layer will induce the residual spin correlation in the second layer. The residual correlation is then presented to the second layer to train the coupling $J_2$ and so on. So the deeper layer should be designed to resolve the residual correlations that can not be resolved in the previous layers. Bearing this physical picture in mind, we propose the following feasible domain 
\eq{\label{eq: J reg}
J_1\geq J_2\geq J_3\geq\cdots\geq 0,}
where $J_z$ is the coupling strength in the $z$th layer. In the  algorithm implementation, the condition \eqnref{eq: J reg} is checked at each training step. If the condition is violated, the parameters $J_z$ will be pulled back to the nearest boundary point of the feasible domain. In the machine learning terminology, \eqnref{eq: J reg} can be considered as a \emph{regularization} that coordinates the training among different layers and effectively prevents overfitting in the first several layers.  

\section{Numerical Results}\label{sec: numerics}

\subsection{Training Set Preparation}

Computing the entanglement entropies for a generic quantum many-body state is difficult. As a proof of concept, we choose the free fermion system to demonstrate the idea of EFL. Consider $N$ copies of the (1+1)D Majorana fermion chain,\cite{Kitaev:2001un} described by the Hamiltonian
\eq{H=\sum_{a=1}^{N}\sum_i\ii \left(1+m(-1)^i\right)\chi_{i,a}\chi_{i+1,a},
}where $\chi_{i,a}$ is the Majorana fermion operator of the flavor $a$ on the site $i$, satisfying $\{\chi_{i,a},\chi_{j,b}\}=\delta_{ij}\delta_{ab}$. The Majorana coupling $(1+m(-1)^i)$ has a staggered pattern along the chain, such that each unit cell contains two sites. $m\in(0,1)$ and $m\in(-1,0)$ correspond to two different gapped topological phases of the fermions, which are separated by the quantum phase transition at $m=0$. The critical point $m=0$, the fermion become gapless and the system is described by a (1+1)D conformal field theory (CFT) with central charge $c=N/2$ (where $N$ is the fermion flavor number). The central charge $c$ and the fermion mass $m$ are two parameters that control the entanglement feature of the Majorana chain. We will tune these two parameters to study their effects on the holographic geometry.

The entanglement entropy of a free fermion state $\ket{\Psi}$ can be efficiently calculated from the fermion correlation function.\cite{Peschel:2003gp,Peschel:2009sv} Let $C_A=\bra{\Psi}\chi\chi^\intercal\ket{\Psi}|_A$ be the fermion correlation restricted to the entanglement subregion $A$, the 2nd Renyi entanglement entropy is then given by
\eq{\label{eq: S fermion}
S_\Psi(A)=-\frac{1}{2}\Tr\ln\big(C_A^2+(1-C_A)^2\big).}
We can then collect the entropy $S_\Psi(A)$ over arbitrary region $A$. The entanglement cut is always placed between the unit cells (i.e.\,the region $A$ always contains complete unit cells). Therefore the local Hilbert space dimension in each unit cell is $D_\partial=2^N=2^{2c}$. Correspondingly, the external pinning field in the Ising model is set by $h=\frac{1}{2}\ln D_\partial=c\ln 2$.

In the following, we will perform the EFL on the ground state of the Majorana fermion chain. The lattice is fixed to the size of 64 sites (i.e.\,32 unit cells) with the periodic boundary condition. The entanglement features are collected from \eqnref{eq: S fermion} and then served to the machine as the training data. For the 32-unit-cell fermion chain, there are altogether $2^{32}$ possible choices of the entanglement region $A$ (as each unit-cell can choose to be included in the region $A$ or not). Obviously, it is both unfeasible and unnecessary to collect $S_\Psi(A)$ for all these $2^{32}$ regions. We will only collect a subset of them. There are several options to choose the sampling ensemble of entanglement regions. 

Option\;(1) is to sample all of them with equal probability. With this sampling scheme, most of the entanglement regions will contains multiple small and disconnected intervals. Consequently, this sampling is not efficient at conveying  large-scale entanglement features for large single intervals (which represent the correlations between far-separated entanglement cuts in the Ising model language). 

Option\;(2) is to sample only single interval regions. As the interval length varies, these regions cover different scales of entanglement features, but the multi-partite entanglement features are missing. We will use this single-interval data for some testing cases to see if the machine has the generalization ability to predict multi-interval entropies form single-interval data. 

Option\;(3), the most comprehensive one, is to weight the entanglement region $A$ by the number of intervals $n_A$ in $A$, such that the probability distribution $p(A)\sim e^{-n_A/\bar{n}}$ is controlled by the average interval number $\bar{n}$. We may tentatively take $\bar{n}=2$, which provides a nice balance between the single-interval and the multi-interval entanglement features. We call this the interval-weighted sampling scheme for the entanglement regions. As we have checked in our numerics, the choice of $\bar{n}$ does not affect training result much (which may be an indication of the internal consistency in entanglement features collected at different interval numbers).

\subsection{Choosing the Central Charge}

We first fix the fermion mass to $m=0$ and run the EFL on the hyperbolic network architecture. The visible layer has 32 units, matching the 32 unit cells of the fermion chain. Each deeper layer halves the number of units, so the number of units per layer vanishes after five layers, and the network can not go deeper. A uniform weight $I_z$ (or equivalently the Ising coupling $J_z=I_z/4$) is assigned to all links in the same layer, where $z=1,2,\cdots,5$ labels the layer depth.

\begin{figure}[htbp]
\begin{center}
\includegraphics[width=0.418\textwidth]{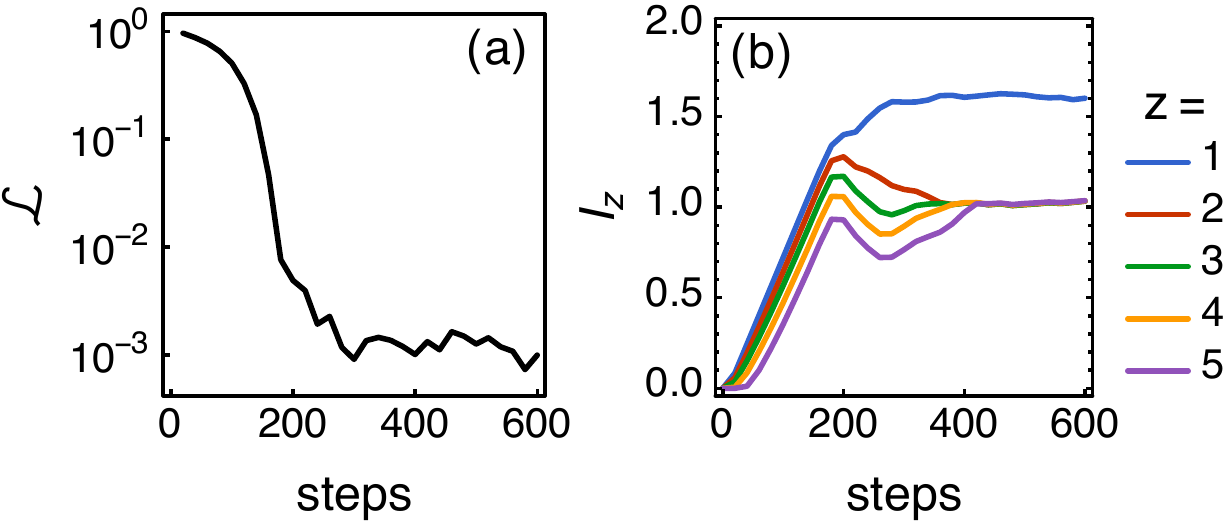}
\caption{Typical training curves of (a) the objective function $\mathcal{L}$ and (b) the edge weight $I_z$ in each layer.}
\label{fig: curve}
\end{center}
\end{figure}

We adopt the supervised learning approach described in \eqnref{eq: L supervised}. The EFL algorithm is implemented\footnote{The source code is available at the GitHub repository \texttt{https://github.com/EverettYou/EFL}.} on the TensorFlow\cite{TF2015} system using Adam\cite{Kingma:2014ev} optimizer. We use the interval-weighted scheme to sample the entanglement regions and prepare the training data for this study. As shown in \figref{fig: curve}(a), the (relative) loss $\mathcal{L}$ decreases with the training steps and converges to $\sim10^{-3}$ eventually. Although the learning algorithm is deterministic, noise is still introduced by the randomly batched training data, leading to the fluctuations of $\mathcal{L}$. Nevertheless, the noise in the training data will not wash out the gradient signals in deep layers, thus the deep network is still trained efficiently.

Driven by the training data, the edge weight $I_z$ develops one layer after another as shown in \figref{fig: curve}(b). Apart from the first layer weight $I_1$, the rest of the weights all converge to the same value controlled by the regularization \eqnref{eq: J reg}. If the regularization condition is lifted, we observe that the machine has the tendency to develop unphysical weights to overfit the data.

\begin{figure}[htbp]
\begin{center}
\includegraphics[width=0.425\textwidth]{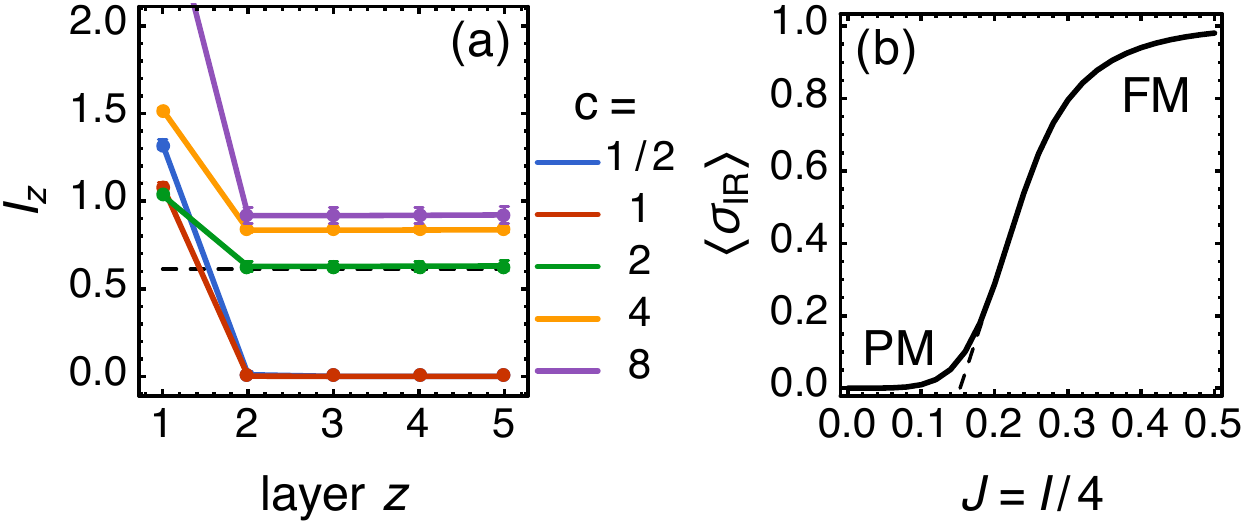}
\caption{(a) Final values of the edge weights $I_z$ on different central charges. The dashed line marks the level of the critical weight $I_c\simeq 0.6$. (b) The IR spin magnetization under UV pinning vs the Ising coupling $J$. The critical $J_c\simeq0.15$ is estimated from extrapolation the activation slope.}
\label{fig: central}
\end{center}
\end{figure}

We take the final values of the weights $I_z$ and plot them in \figref{fig: central}(a). As we tune the central charge $c$ of the fermion chain, the behavior of $I_z$ undergoes a transition around $c=2$. When the central charge is smaller than that (e.g.\,$c=1/2,1$), the deep layers will not be trained. This corresponds to an order-disorder transition of the Boltzmann machine. Smaller central charge means weaker entanglement and smaller edge mutual information in the RTN. Since the edge mutual information $I$ maps to the Ising coupling $J=I/4$, decreasing the coupling $J$ could drive the system into the paramagnetic phase. Then the original assumption on the large edge mutual information fails and the physical picture of representing the entanglement entropy by the domain wall energy in the holographic bulk no longer holds. To estimate the critical coupling $J_c$ on the hyperbolic network, we first pin the boundary spins to the same direction and then measure the magnetization of the spin at the deepest layer to see if the magnetization can propagate through the system all the way from the boundary to the deepest layer in the bulk. As shown in \figref{fig: central}(b), we found an activation behavior in the magnetization curve, which roughly divides the coupling $J$ into paramagnetic-like or ferromagnetic-like regimes. Although the transition is smeared out in the finite-sized system, we can still give an estimate of the critical $J_c\simeq 0.15$ (or $I_c\simeq 0.6$) from the extrapolation of the activation slope. \figref{fig: central}(a) indeed shows that as $I_z$ drops below the level of $I_c$, the training signal disappears and the deep layer weights cease to develop.

In the AdS/CFT duality, the central charge $c$ of a holographic CFT$_2$ is universally given by $c=3\ell/2G_N$,\cite{Brown:1986qi,Ryu:2006fj} where $\ell$ is the AdS radius and $G_N$ is the Newton constant in three dimensional gravity. Our approach of fixing the tensor network architecture and training the edge weights corresponds to fixing the AdS radius. Then changing the central charge $c$ effectively changes the gravitational constant $G_N$. Large $c$ corresponds to small $G_N$ and hence weakly coupled classical gravity. The classical holographic geometry can be represented by the classical network geometry that can be trained by the EFL. As the central charge $c$ gets small, the gravity crosses over from classical to quantum and the EFL ceases to produce a sensible result. Therefore, in the following, we will fix the central charge at $c=4$ on the classical side.

\subsection{Single-Interval Entanglement Entropy}

\begin{figure}[htbp]
\begin{center}
\includegraphics[width=0.44\textwidth]{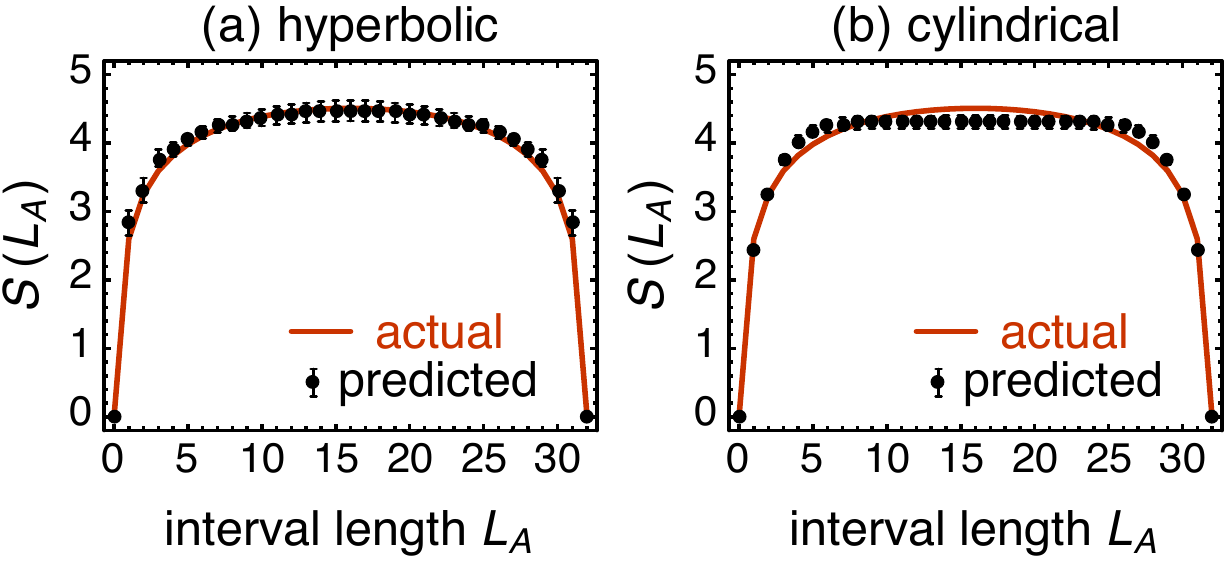}
\caption{Fitting the single-interval 2nd Renyi entropy using the machine trained on (a) the hyperbolic and (b) the cylindrical network architecture. The lattice contains 32 unit cells and interval length $L_A$ is measured in unit cells.}
\label{fig: single}
\end{center}
\end{figure}

For the critical fermion chain $m=0$, it is known that the single-interval Renyi entanglement entropy (i.e.\,the entanglement region $A$ is a single continuous interval) follows the logarithmic law $S(L_A)\sim\ln L_A$ in the thermodynamic limit.\cite{Calabrese:2004ve} To see how well the RTN can reproduce this logarithmic entropy scaling after training, for this study we serve the machine with only the single-interval 2nd Renyi entanglement entropies taken from a critical fermion chain of 32 unit cells (calculated from \eqnref{eq: S fermion} using the lattice model). After the training, we ask the machine to reproduce the entanglement entropies over the trained intervals and compare the predictions with the actual values. The result is shown in \figref{fig: single}.

On the hyperbolic architecture, the trained model provides a good fitting as in \figref{fig: single}(a). However, on the cylindrical architecture, the fitting gets worse and the regression error is larger as in \figref{fig: single}(b). This is because the expression power of the cylindrical network is not strong enough to capture the logarithmic entropy scaling. Naively, one may imagine to mimic the hyperbolic geometry on the cylindrical network with the weights that gradually decay with the layer depth. However the problem is that as the edge weight (Ising coupling) gets smaller than the critical value, the deeper layers will enter the paramagnetic phase and lose the learning signal. As a consequence, only the first several layers will be trained in the cylindrical network typically, which results in an area law entangled RTN structure (similar to MPS).\footnote{Another idea to realize the hyperbolic geometry on the cylindrical network is to allow the translation symmetry breaking, so that some bonds in the deep layers can be turned off in a pattern similar to the hyperbolic network. In this way, it is possible to realize the logarithmic entropy scaling on a cylindrical network. We will leave this possibility for future research.} Therefore when the CFT entanglement feature is fed to the cylindrical Boltzmann machine, the machine will try to fit the entropy data with an area law curve, which can be seen from the flat top behavior of the prediction curve in \figref{fig: single}(b). Thus for CFT states, the hyperbolic network generally provides a better fit to the logarithmic entropy scaling compares to the cylindrical network. It is conceivable that if the machine is allowed to adjust its architecture during the training, the EFL will generate a training signal to drive the cylindrical network towards the hyperbolic network for the CFT states. However, dynamically updating network architectures in the training process is still technically challenging, we will leave this possibility for future study.

\subsection{Multi-Interval Entanglement Entropy}

\begin{figure}[htbp]
\begin{center}
\includegraphics[width=0.22\textwidth]{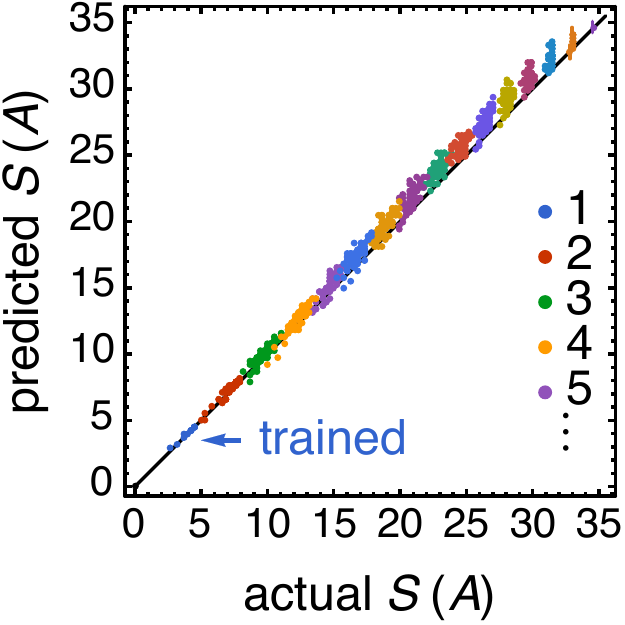}
\caption{Predicted vs actual entropy over multi-interval entanglement regions. Each pair is classified by the interval number in color. Only the single-interval data was trained.}
\label{fig: multi}
\end{center}
\end{figure}

To test the prediction power of the RTN model, we train a hyperbolic network using single-interval entanglement entropies and ask if the network can predict multi-interval entanglement entropies. Let us use different colors to label the entanglement entropies over different numbers of intervals, and plot the predicted entropy against the actual entropy in \figref{fig: multi}. In the training phase, only the single-interval data is presented to the machine. After the training, the machine was able to predict multi-interval entanglement entropies, which was not in the training set. If the prediction is perfect, then all the points should fall along the diagonal line in \figref{fig: multi}. We can see the points do line up nicely, especially when the number of intervals is small. The overall prediction accuracy is $\sim95\%$. 

This demonstrates the prediction power of the RTN model. However, this may not be very surprising. Since the multi-interval entanglement entropy is related to the single-interval ones
\eq{S(A\cup B)=S(A)+S(B)-I(A, B).} 
If the mutual information $I(A, B)$ is small, the multi-interval entropy is dominated by the additive part $S(A\cup B)\simeq S(A)+S(B)$, which is relatively easy to capture. So we will turn to the sub-additive part (i.e.\,the mutual information) in the following.

\subsection{Mutual Information}

\begin{figure}[htbp]
\begin{center}
\includegraphics[width=0.414\textwidth]{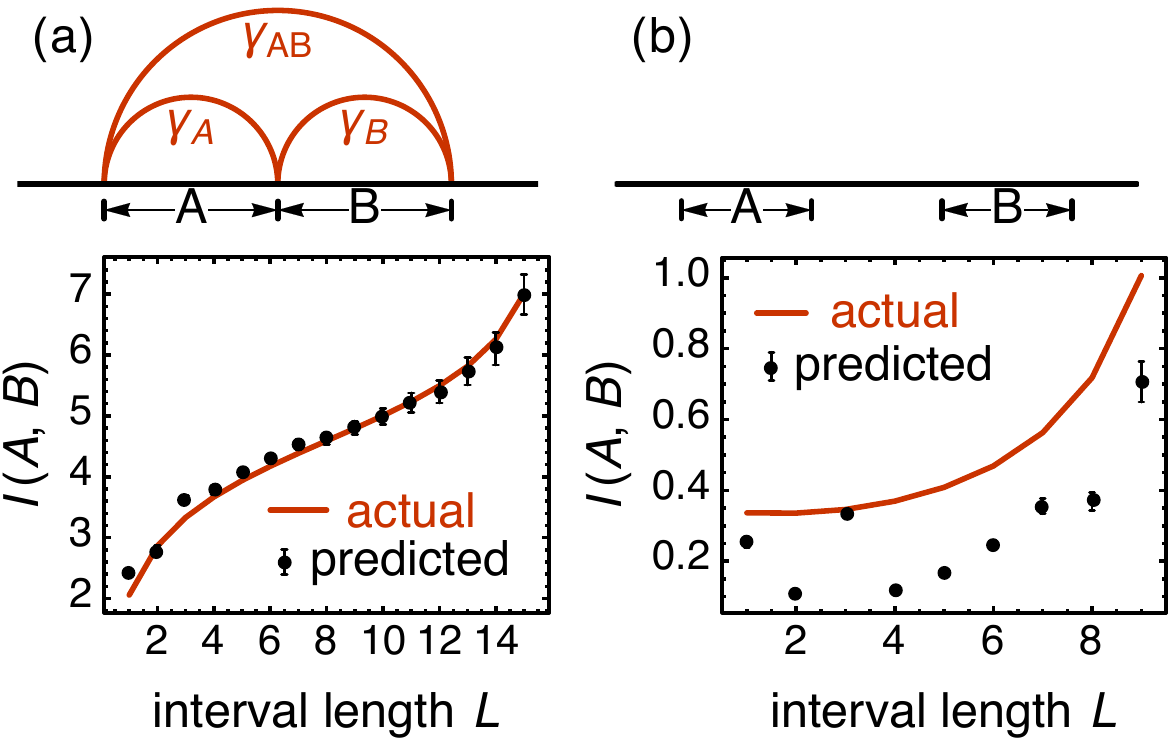}
\caption{Fitting the mutual information of (a) two adjacent equal-length intervals and (b) two separated equal-length intervals with the separation region of the same length as the interval length.}
\label{fig: mutual}
\end{center}
\end{figure}

We found that for adjacent intervals, the RTN model can still fit the mutual information well, as shown in \figref{fig: mutual}(a). There is a geometric interpretation of this type of mutual information in the holographic bulk. According to the Ryu-Takayanagi formula,\cite{Ryu:2006fj} the entanglement entropy $S(A)$ of the interval $A$ is proportional to the area of the minimal surface $\gamma_A$, which, in the AdS$_3$ space, is also the geodesic line connecting the two boundary points of the interval $A$. Therefore the mutual information of adjacent intervals $A$ and $B$ corresponds to
\eqs{I(A,B)&=S(A)+S(B)-S(AB)\\
&=\frac{1}{4G_N}(|\gamma_{A}|+|\gamma_{B}|-|\gamma_{AB}|).}
$\gamma_{A}$, $\gamma_{B}$ and $\gamma_{AB}$ form the three sides of a triangle in the holographic bulk. The mutual information measures how much is the sum of the two sides greater than the third side. This indicates that the machine gets a grasp of the holographic geometry in its neural network, so it can provide a good prediction of the mutual information that has classical geometric interpretations.

In contrast, for separated intervals, the predicted mutual information is obviously less than the actual value by quite a large amount, as shown in \figref{fig: mutual}(b). This is actually not a problem of our algorithm, but has a deep physical origin. States with semi-classical dual which satisfies RT formula are necessarily strongly correlated and contain a lot of multi-partite entanglement. For example, it is known that holographic states have large and negative tripartite information,\cite{Hayden:2016zm,Hosur:2016te} in contrast from the free fermion theory. RTN is designed to describe holographic states, which have much smaller mutual information between separated intervals compared to that in the free fermions. The free fermion conformal field theory has many low-dimension operators, which corresponds to light matter fields in the dual gravity theory. In our approach, these matter field fluctuations are not taken into account, which partially explains the  the deficit of mutual information in \figref{fig: mutual}(b). Also, our approach only captures the optical classical geometry and does not include the quantum fluctuation of geometries around the classical saddle points. How to go beyond the planar graph EFL and include the fluctuation effect of both matter fields and geometries is an interesting topic for future research.

\subsection{Emergent Holographic Geometry}

\begin{figure}[htbp]
\begin{center}
\includegraphics[width=0.473\textwidth]{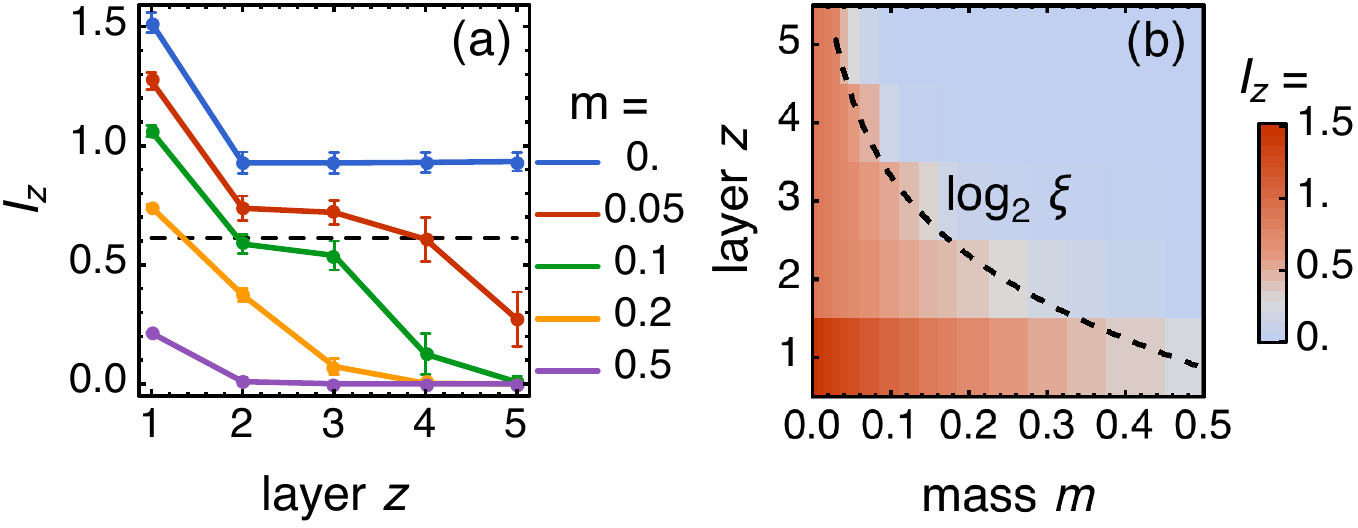}
\caption{(a) The edge weight $I_z$ in each layer, trained from the fermion model with different mass $m$. (b) Color plot of the edge weight $I_z$ as a function of mass $m$ and the layer depth $z$. The dashed line is the curve of $\log_2\xi$ with $\xi$ taken from \eqnref{eq: xi}.}
\label{fig: mass}
\end{center}
\end{figure}

Finally, we turn on the fermion mass $m$. The fermion correlation length $\xi$ becomes finite and is given by
\eq{\label{eq: xi}\xi^{-1}=\frac{1}{2}\ln\frac{1+|m|}{1-|m|}.}
In the holographic bulk, the fermion mass caps off the IR region at the scale $z_\text{IR}\sim\ln\xi$. Because the entanglements are resolved in the UV layers of the RTN at this scale, the network ceases to grow deeper and the holographic space ends. As the mass $m$ is turned on,  the edge weight will start to fade away from the deepest layer,  as shown in \figref{fig: mass}(a). With increasing mass, the fade-off scale $z_\text{IR}$ moves from IR (large $z$) toward UV (small $z$), see \figref{fig: mass}(a).

We scan over a range of mass $m\in[0,0.5]$. At each $m$, we train the machine and obtain the edge weight $I_z$. The result is shown in \figref{fig: mass}(b). There is a clear boundary where the holographic space terminates. This boundary matches the theoretical expectation $z_\text{IR}=\log_2\xi$ nicely (we take $\log_2$ here because of each deeper layer halves the number of unit in the hyperbolic network architecture). This demonstrates how the AdS$_3$ spacial geometry emerges as we gradually decrease the mass $m$ and drive the boundary system toward the CFT$_2$.

\section{Discussions and Summary}

In this work, we have restricted the entanglement feature to the 2nd Renyi entropies. It is actually conceptually more natural to include all orders of Renyi entropies over all regions in the entanglement feature.\footnote{More generally, one can consider all local unitary (LU) invariants, which are of the form ${\rm tr}\left[\rho^{\otimes k}\prod_{i}g_i\right]$ with $g_i$ on each site $i$ an arbitrary component of the $S^k$ permutation group.} 
Ref.\,\onlinecite{Hayden:2016zm} shows that the $n$th Renyi entropy of the RTN state can be mapped to the free energy difference of an $S_n$ model in the large bond dimension limit. In the $S_n$ model, each vertex $v\in\scV$ hosts a permutation group element $\sigma_v\in S_n$, coupled together via the energy functional
\eq{E_\scG[\sigma,\tau]=-\sum_{e\in\scE}\chi_e\Big(\prod_{v\in\partial e}\sigma_v\Big)-\sum_{v\in\scV_\partial}\chi_\partial(\tau_v^{-1}\sigma_v),}
where $\chi_e(g)$ and $\chi_\partial(g)$ are class functions that only depend on the cycle type $l_g^\alpha$ of the permutation $g$ (i.e.\,$l_g^\alpha$ is the length of the $\alpha$th cycle in $g$). More specifically, we have
\eqs{\chi_e(g)&=\frac{1}{2}\sum_\alpha(l_g^\alpha-1)I_e^{(l_g^\alpha)},\\
\chi_\partial(g)&=\sum_\alpha(l_g^\alpha-1)\ln D_\partial.}
The $\chi_e$ function is parameterized by the edge mutual information $I_e^{(l)}$ for Renyi index $l=2,\cdots,n$. The $\chi_\partial$ term describes the boundary pinning field that pins the boundary configuration to another set of permutation group elements $\tau_v\in S_n$. By defining $e^{-F_\scG[\tau]}=\sum_{[\sigma]}e^{-E_\scG[\sigma,\tau]}$, we can consider $S_\scG[\tau]=F_\scG[\tau]-F_\scG[\tau=1]$. If we allow $\tau_v$ to take all group elements in $S_n$ (not limited to the cyclic permutations), the function $S_\scG[\tau]$ actually includes the RTN entanglement entropies over all regions for all Renyi index up to $n$. In principle, using the $S_n$ model, entanglement entropies of different Renyi indices (up to $n$) can all be put together as the training data for the EFL, and the edge mutual information of all Renyi indices (up to $n$) will be trained simultaneously. However,  the efficient training method for Boltzmann machines on $S_n$ models is still lacking, so the above idea is still not practical yet. 

Despite the technical difficulties, the philosophy behind EFL is clear. For a quantum many-body state with a given tensor factorization of the Hilbert space (which specify the ``real space basis"), one can forget about operator-specific information such as particular correlation functions, and focus on the local unitary invariant information. All local unitary invariant properties of the wavefunction can be considered as entanglement features of the wavefunction, which include the bipartite entanglement properties and also many more multipartite entanglement properties. From the point of view of gravitational dual, it is interesting to make an analog with the black hole no-hair theorem\cite{Misner:1973sb,Israel:1967kr,Israel:1968sv,Carter:1971sw}. The non-invariant features are removed and the geometry only encodes the local unitary invariant features, in the same way how the area of the black hole is proportional to its entropy and is independent from details of the initial state. The random average in RTN serves as a technical tool to remove ``hairs" of a many-body state, where the operator specific information is erased by the random tensor, leaving only the entanglement features encoded in the network structure. 
Consequently, RTN can be potentially a useful framework for characterizing other phenomena in which entanglement features play an essential role, such as the many-body localization-thermalization transition,\cite{Vasseur:2017om} which is essentially a transition about entanglement structures. The EFL provides us an approach to construct the RTN and to optimize its structures, which could be a useful tool for the study of quantum chaotic dynamics and localization/thermalization.

In summary, the goal of the EFL is to construct an optimal RTN state that best fits the entanglement properties of a given quantum many-body state. The problem similar to the task of feature learning, which extracts the features hidden in the training data and encode them into the structure of the neural network. This analogy is made concrete by mapping the RTN to the Boltzmann machine and train the machine with the entanglement entropies over all subregions. As the entanglement feature is learned, the machine develops a neural network, whose network geometry can be interpreted as the emergent holographic geometry of the given quantum many-body state.

\begin{acknowledgements}
The authors would like to acknowledge the helpful discussions with Pan Zhang, Yanran Li, and Roger Melko. YZY also benefits from the 2016 and 2017 Swarma Club Workshop on ``Geometry, Complex Network and Machine Learning'' sponsored by Kai Feng Foundation. This work is supported by the National Science Foundation through the grant No. DMR-1151786 (ZY), and the David and Lucile Packard Foundation (XLQ).

\end{acknowledgements}

\appendix
\section{Entanglement Entropy of RTN States}\label{sec: RTN}
The RTN entanglement entropy defined in \eqnref{eq: S def} can be equivalently expressed as
\eq{\label{eq: S op}e^{-S_{\scG}(A)}=\dsE \frac{\Tr(\ket{\scG}\bra{\scG})^{\otimes 2}\hat{\tau}(A)}{\Tr (\ket{\scG}\bra{\scG})^{\otimes 2}},}
where $\dsE$ denotes the average over the RTN ensemble, and $\hat{\tau}(A)$ is the swap operator in the subregion $A$. It can be factorized to each boundary vertex as
\eq{\label{eq: tau A op}
\hat{\tau}(A)=\bigotimes_{v\in\scV_\partial}\hat{\tau}_v,\quad \hat{\tau}_v=\left\{\begin{array}{ll}\dia{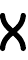}{12}{-3} & v\in A, \\ \dia{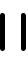}{12}{-3} & v\in \bar{A}.\end{array}\right.}
The operator $\hat{\tau}_v$ swaps the replicated local Hilbert space $(\mathcal{H}_v^\text{phy})^{\otimes2}$ on the vertex $v\in A$, otherwise it is an identity operator acting on the vertex $v\in\bar{A}$.

To evaluate \eqnref{eq: S op}, we first introduce the rules for the ensemble average of the random state. Suppose $\ket{\psi}$ is a random state in an $N$-dimensional Hilbert space, because the random state ensemble is $\mathrm{SU}(N)$ symmetric, due to the Schur's lama, the ensemble average of $\ket{\psi}\bra{\psi}$ must be proportional to the identity matrix in respect of the $\mathrm{SU}(N)$ symmetry. Under appropriate normalization, we can set $\dsE \ket{\psi}\bra{\psi}=\mathbf{1}$. Introducing the graphical representation of the random state,
\eq{\ket{\psi}=\dia{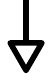}{12.5}{-4}, \bra{\psi}=\dia{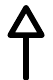}{12.5}{-4},}
the formula $\dsE \ket{\psi}\bra{\psi}=\mathbf{1}$ can be represented as
\eq{\dsE\dia{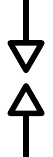}{25}{-8}=\dia{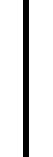}{25}{-8}.}
For duplicated case, the formula is generalized to
\eq{\label{eq: E rho2}\dsE\Big(\dia{braket}{25}{-8}\Big)^{\otimes 2}=\dsE\dia{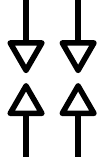}{25}{-8}=\dia{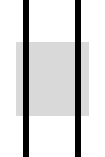}{25}{-8}+\dia{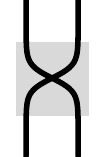}{25}{-8}=\sum_{\sigma\in S_2}\dia{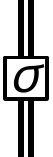}{25}{-8},}
or $\dsE(\ket{\psi}\bra{\psi})^{\otimes 2}=\sum_{\sigma\in S_2}\hat{\sigma}$, as a result of the $\mathrm{SU}(N)\times S_2$ symmetry.

Consider a RTN state on a graph with two vertices, each connected to an external edge (physical leg).
\eq{\ket{\scG}=\dia{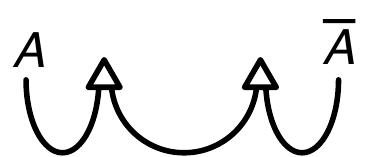}{25}{-8}.}
Assuming the left and the right external edges are ascribed to the entanglement regions $A$ and $\bar{A}$ respectively. Using \eqnref{eq: E rho2}, one can evaluate
\eqs{\label{eq: E rho2tau}&\dsE\Tr(\ket{\scG}\bra{\scG})^{\otimes 2}\hat{\tau}(A)\\
=&\dsE\dia{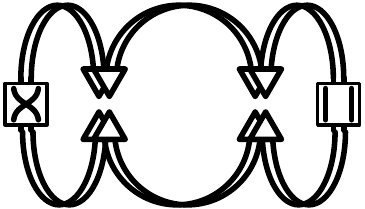}{34}{-14}=\sum_{[\sigma]}\dia{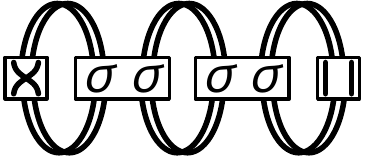}{26}{-12}\\
=&\sum_{[\hat{\sigma}]}w(\dia{X}{12}{-3},\hat{\sigma}_1)w(\hat{\sigma}_1,\hat{\sigma}_2)w(\hat{\sigma}_2,\dia{I}{12}{-3}),}
where $\hat{\sigma}_i\in S_2$ arise from the ensemble average of the random states in the bulk. The weight function $w(\hat{\sigma}_1,\hat{\sigma}_2)$ is actually a function of $\hat{\sigma}_1^{-1}\hat{\sigma}_2$, which can be expressed in terms of the 2nd Renyi mutual information $I_e$ of the entangled pair state along the edge 
\eq{w(\hat{\sigma}_1,\hat{\sigma}_2)=\left\{\begin{array}{ll}1&\text{if }\hat{\sigma}_1^{-1}\hat{\sigma}_2=\dia{I}{12}{-3},\\e^{-I_e/2}&\text{if }\hat{\sigma}_1^{-1}\hat{\sigma}_2=\dia{X}{12}{-3}.\end{array}\right.}
If we represent the $S_2$ variable $\hat{\sigma}_i$ by the Ising variable $\sigma_i=\pm1$, the weight function has a more compact form $w(\sigma_1,\sigma_2)\propto e^{-J_e\sigma_1^{-1}\sigma_2}$ where $J_e=I_e/4$. For external edges, $J_e$ is replaced by $h=\frac{1}{2}\ln D_\partial$ where $D_\partial$ is the boundary bond dimension. So \eqnref{eq: E rho2tau} can be map to the partition function of an Ising model with fixed boundary condition $\tau(A)$, 
\eq{\dsE\Tr(\ket{\scG}\bra{\scG})^{\otimes 2}\hat{\tau}(A)=e^{-F_\scG[\tau(A)]},}
where $e^{-F_\scG[\tau]}=\sum_{[\sigma]}e^{-E_\scG[\sigma,\tau]}$ and
\eq{E_\scG[\sigma,\tau]=-h\tau_1\sigma_1-J_{12}\sigma_1\sigma_2-h\tau_2\sigma_2.}
It is straightforward to generalize the energy functional to generic graphs, given in \eqnref{eq: E Ising}. Correspondingly, \eqnref{eq: tau A} is just a rewritten of \eqnref{eq: tau A op} in terms of the Ising variables.

In the large bond dimension limit (large $I_e$), the ensemble average of the fraction in \eqnref{eq: S op} can be approximated by average of the numerator and the denominator separately.
\eqs{e^{-S_\scG(A)}\simeq&\frac{\dsE\Tr(\ket{\scG}\bra{\scG})^{\otimes 2}\hat{\tau}(A)}{\dsE\Tr (\ket{\scG}\bra{\scG})^{\otimes 2}}\\=&e^{-F[\tau(A)]+F[\tau(\emptyset)]}.}
Hence we have arrived at $S_\scG(A)\simeq F[\tau(A)]-F[\tau(\emptyset)]$, verifying the result in \eqnref{eq: S RTN}. Ref.\,\onlinecite{Hayden:2016zm} has shown that the approximation of distributing the ensemble average into the fraction is valid in the large $I_e$ limit by analyzing the fluctuation. A more careful treatment away from the that limit is provided in Ref.\,\onlinecite{Vasseur:2017om}.

\section{Planar Graph Ising Model}\label{sec: Ising}

In this appendix, we will review the systematic approach to calculate the free energy $F$ of the Ising model on a planar graph $\scG=(\scV,\scE)$, following Ref.\,\onlinecite{Fisher:1961il,Kasteleyn:1961bl}.
\eq{\label{eq: Z Ising}Z=e^{-F}=\sum_{[\sigma]}e^{-E[\sigma]},\quad E[\sigma]=-\sum_{e\in\scE}J_e\prod_{v\in\partial e}\sigma_v.}
First of all, every planar graph can be triangulated by adding virtual edges, across with the Ising coupling $J_e=0$ is simple zero. If the boundary spins are also coupled to external Zeeman field $h_v$, one can consider introducing a fictitious spin at infinity and coupling all the boundary spins to the fictitious spin with the coupling strength set by $h_v$. This effective doubles the system by its $\mathbb{Z}_2$ symmetry (the Ising spin flip symmetry) counterpart, which only brings a factor 2 to the partition function but does not affect the free energy calculation. With the tricks of virtual edges and the fictitious spin, we only need to consider the $\mathbb{Z}_2$ symmetric Ising model on the triangulated planar graph.

\begin{figure}[htbp]
\begin{center}
\includegraphics[width=0.36\textwidth]{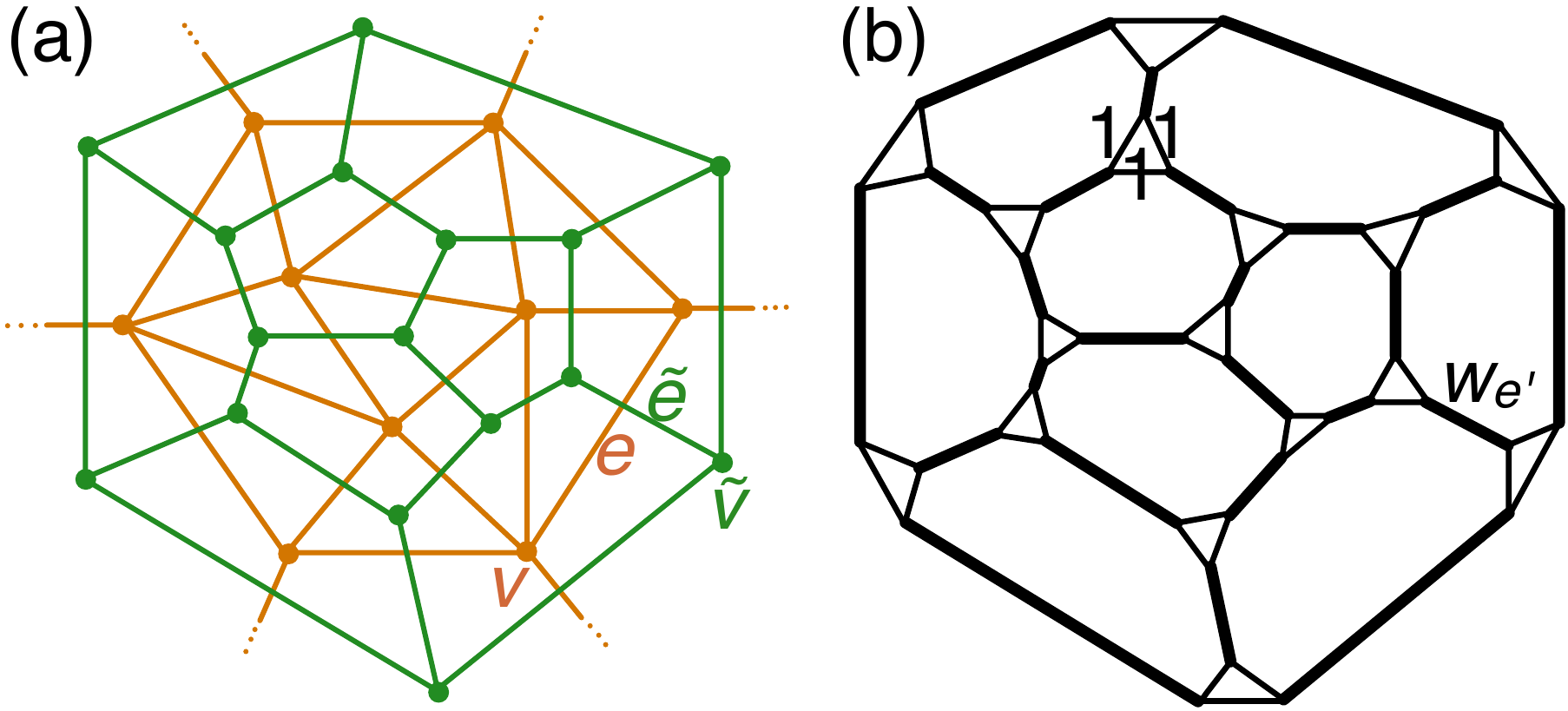}
\caption{(a) The original graph (in orange) and its dual graph (in green). Each edge $e$ in the original graph is dual to a unique edge $\tilde{e}$ in the dual graph, such that $e$ and $\tilde{e}$ intersect. (b) The extended graph (star lattice) by expanding each site to a three sites in a triangle.}
\label{fig: dual}
\end{center}
\end{figure}

Every triangulated planar graph has a dual trivalent graph $\tilde{\scG}=(\tilde{\scV}, \tilde{\scE})$, as shown in \figref{fig: dual}(a), on which the Ising model is mapped to a loop model. Each Ising domain wall is interpreted as a loop on the dual lattice. Introduce the $\mathbb{Z}_2$ variable $l_{\tilde{e}}$ on the dual edges $\tilde{e}$, such that $l_{\tilde{e}}=1$ corresponds to a loop through the edge $\tilde{e}$ and $l_{\tilde{e}}=0$ corresponds to no loop. The partition function \eqnref{eq: Z Ising} can be mapped to\cite{Globerson:2007zr}
\eqs{\label{eq: Z loop}
Z&=\sum_{[\sigma]}\prod_{e\in\scE}e^{J_e\prod_{v\in\partial e}\sigma_v}\\=&Z_0\sum_{[l]}\prod_{\tilde{e}\in\tilde{\scE}}w_{\tilde{e}}^{1-l_{\tilde{e}}}\prod_{\tilde{v}\in\tilde{\scV}}\delta_{\mathbb{Z}_2}\Big(\sum_{\tilde{e}\in\dd\tilde{v}}l_{\tilde{e}}\Big),}
where the weight is $w_{\tilde{e}}=e^{2J_{e}}$ (where $e$ the edge in the original graph that is dual to the edge $\tilde{e}$ in the dual graph) and the factor $Z_0=e^{-F_0}$ is given by $F_0=\sum_{e\in\scE}J_e$. The delta function $\delta_{\mathbb{Z}_2}$ over the $\mathbb{Z}_2$ group imposes the close loop constraint. Unlike conventional loop models, here each segment of the loop (the domain wall) is given a trivial weight $1$, while the edge without the loop is given a greater weight $w_{\tilde{e}}\geq 1$ (for $J_e\geq0$) instead. In this way the domain wall is still relatively suppressed in the partition function. The overall factors generated in this weight rescaling are all absorbed into $Z_0$.

Further expanding each trivalent site into a triangle, as shown in \figref{fig: dual}(b), the loop model can be mapped to a dimmer model,\cite{Globerson:2007zr} where the loop configuration is replaced by the transition graph of dimmer configurations. Let $\Omega$ be the set of all dimmer coverings (perfect matchings) of the extended graph $\scG'$ in \figref{fig: dual}(b), the partition function \eqnref{eq: Z loop} becomes 
\eq{\label{eq: Z dimmer}
Z=Z_0\sum_{M\in\Omega}\prod_{e'\in M}w_{e'}}
In the dimmer model, each thick edge covered by the dimmer is weighted by $w_{e'}=e^{2J_{e}}$. The remaining thin edges all share $w_{e'}= 1$.

\begin{figure}[htbp]
\begin{center}
\includegraphics[width=0.36\textwidth]{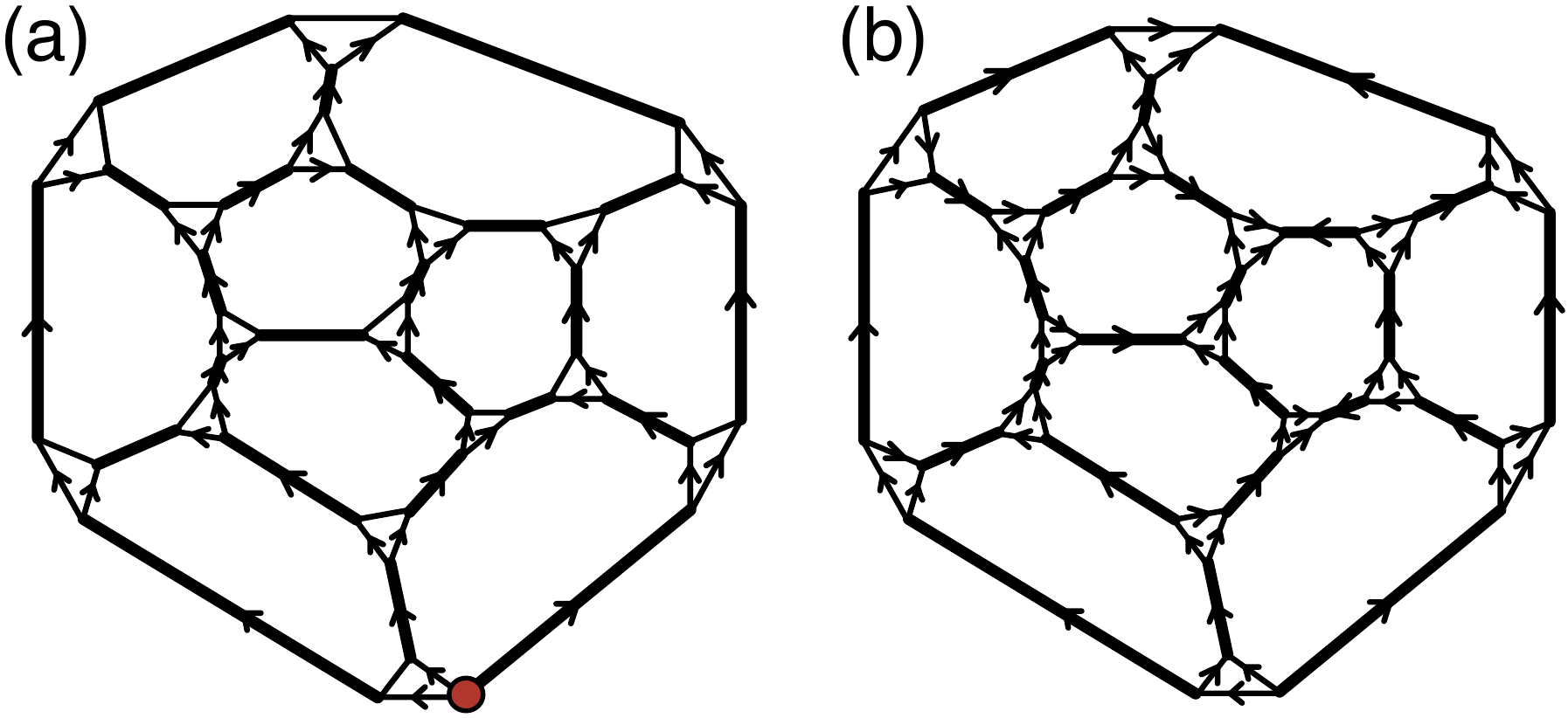}
\caption{Systematic assignment of the Kasteleyn orientation on planar graph. (a) Start from an arbitrary vertex (mark by the red dot) and build a spanning tree. (b) Close the loops respecting the clockwise-odd rule.}
\label{fig: orient}
\end{center}
\end{figure}

The partition function of the dimmer model \eqnref{eq: Z dimmer} can be formulated as a path integral of free Majorana fermions, with fermion spin structure specified by the Kasteleyn orientation.\cite{Kasteleyn:1961bl,Temperley:1961dq} The insight is that every non-zero term in the Majorana fermion path integral corresponds to a perfect matching on the graph $\scG'$ (on which the dimmer model is defined). To place the fermion system on the graph $\scG'$, each edge must be assigned an orientation, such that for every face (except possibly the external face) the number of edges on its perimeter oriented in a clockwise manner is odd, known as the clockwise-odd rule. Any orientation satisfying the clockwise-odd rule is a Kasteleyn orientation, which ensures all dimmer configurations to be mapped to even fermion parity states. The Kasteleyn orientation can be assigned systematically on planar graphs by first choosing an arbitrary vertex in the graph and build a spanning tree from that vertex, then closing the loops respecting the clockwise-odd rule, as demonstrated in \figref{fig: orient}.

With the Kasteleyn orientation assigned, we can construct the weighted adjacency matrix $A$ of the graph $\scG'=(\scV',\scE')$, such that $\forall i,j\in\scV'$: $A_{ij}=0$ if $\langle ij\rangle$ is not an edge in $\scE'$, $A_{ij}=w_{ij}$ if the orientation on edge $\langle ij\rangle$ runs from $i$ to $j$, and $A_{ij}=-w_{ij}$ otherwise. The partition function can then be shown to be
\eq{Z=Z_0\int\mathcal{D}[\chi] e^{-\frac{1}{2}\chi^\intercal A\chi}=Z_0\pf A.}
So the free energy of the Ising model can be calculated from
\eq{F=F_0-\ln\pf A,}
where $F_0=\sum_{e\in\scE}J_e$ and $A$ is the adjacency matrix of the Kasteleyn oriented extended dual graph $\scG'$.

\bibliography{RTN}
\bibliographystyle{apsrev}
\end{document}